\documentclass[12pt,a4paper]{article}
\usepackage{epsfig}
\newcommand{\bea}{\begin{eqnarray}}
\newcommand{\eea}{\end{eqnarray}}
\begin{document}
\begin{titlepage}
\begin{center}
{\hbox to\hsize{\hfill KEK-TH-890}}

\vspace{4\baselineskip}

\textbf{\Large 
        Neutrino Magnetic Moments and \\ 
\bigskip
        Minimal Supersymmetric SO(10) Model}
\bigskip
\bigskip
\vspace{2\baselineskip}

\textbf{Takeshi Fukuyama%
\footnote{E-Mail: fukuyama@se.ritsumei.ac.jp}
and Tatsuru Kikuchi%
\footnote{E-Mail: rp009979@se.ritsumei.ac.jp }
} \\ 
\bigskip
\textit{\small 
Department of Physics, Ritsumeikan University, Kusatsu, 
Shiga 525-8577, Japan 
}
\vspace{2\baselineskip}

\textbf{Nobuchika Okada%
\footnote{E-Mail: okadan@post.kek.jp} }\\
\bigskip
\textit{\small
Theory Group, KEK, Oho 1-1, Tsukuba, Ibaraki 305-0801, Japan}

\vspace{3\baselineskip}

\textbf{Abstract}\\
\end{center}
\noindent
We examine supersymmetric contributions 
 to transition magnetic moments of Majorana neutrinos.  
We first give the general formula for it. 
In concrete evaluations, 
 informations of neutrino mass matrix elements  
 including CP phases are necessary. 
Using unambiguously determined neutrino mass matrices 
 in recently proposed minimal supersymmetric SO(10) model, 
 the transition magnetic moments are calculated. 
The resultant neutrino magnetic moments 
 with the input soft supersymmetry breaking masses 
 being of order 1 TeV
 are found to be roughly an order of magnitude larger 
 than those calculated in the standard model extended 
 to incorporate the see-saw mechanism. 
\end{titlepage}
%
\setcounter{footnote}{0}
\newpage
%

The stellar objects like the Sun, White Dwarfs, Supernovae, 
 and Magnetors have magnetic field 
 roughly from $10^3$ to $10^{15}$ Gauss (G).  
The motions of neutrinos in the medium of these objects 
 are affected by these magnetic fields 
 if neutrinos have the sufficient magnitude of magnetic moments. 
Experiments show the following limits on the magnetic moments 
 \cite{PDG}, 
\bea
 \mu_{\nu_e} &\leq& 1.5 \times 10^{-10} \mu_{B},    \nonumber\\
 \mu_{\nu_{\mu}} &\leq& 6.8 \times 10^{-10} \mu_{B},  \nonumber\\
 \mu_{\nu_{\tau}} &\leq& 3.9 \times 10^{-7} \mu_{B} 
\eea
 in units of the Bohr magneton, $\mu_B\equiv \frac{e\hbar}{2m_e}$. 
Astrophysical observations provide severer 
 constraints: 
from anomalous stellar cooling due to the plasmon decay 
 $\gamma\rightarrow \nu\bar{\nu}$ \cite{raffelt},
\bea
 (\sum_{i,j}|\mu_{ij}|^2)^{1/2}\le 3\times 10^{-12} \mu_B,
 \label{plasmon}
\eea
$\nu_L\rightarrow \nu_R$ conversion process 
in SN1987A \cite{ayala} gives 
\bea
\mu_{\nu_e} < (1-4)\times 10^{-12} \mu_B,
\eea
and 
\bea
\mu_{\nu_e}\le 3.9\times 10^{-12} \mu_B
\eea
from solar neutrino observations \cite{lujan} 
for the moderate solar magnetic field $B=50$kG etc. 
Eq.~(\ref{plasmon}) is especially important constraint 
 since it includes all the off-diagonal entries of 
 neutrino magnetic moments. 
If $\mu_{i j}$ is large enough 
 (but is still satisfied by current upper bound), 
\bea 
 \mu_\nu \geq 10^{-14}\mu_B , 
\eea
the neutrino spin flavor precession (SFP) can occur 
 in the solar \cite{akhmedov1} or 
 in the supernovae \cite{akhmedov2} 
 and we can expect the observable effects. 
The comprehensive arguments under the assumption 
 of large neutrino magnetic moment are given, for instance, 
 in the text by Raffelt \cite{raffelt1}. 

In theoretical point of view, it is interesting 
 to examine how large neutrino magnetic moments 
 are obtained in massive neutrino models. 
For the Dirac neutrinos in the standard model 
 extended to include right-handed neutrinos, 
the magnetic moments are found to be \cite{fujikawa}
\bea
\mu_\nu = \frac{3e G_F m_\nu}{8\sqrt{2}\pi^2} 
 \sim 3.20 \times10^{-19} 
 \left(\frac{m_\nu}{1 \mathrm{eV}} \right) \;  \mu_B , 
\label{stdmu}
\eea
which is strongly suppressed 
 owing to the chiral symmetry and the GIM cancellations. 
If neutrinos are Majorana particles, 
 they can only have transition magnetic moments (TMMs)
 because of CPT invariance of the theory. 
The transition matrix corresponding to TMM 
 such as $\nu_i (p) \rightarrow \nu_j(p-q) + \gamma (q) 
 \; (i \neq j) $ is described as  
\bea 
T_{ji} = -i \epsilon^{\mu} \bar{\nu}_j (p-q) 
\left[ 
 2 i \Im F_2(q^2)_{ji} \right] \sigma_{\mu \nu} 
q^\nu \nu_i(p)  \; , 
\eea 
where $F_2(q^2)_{ji}$ is the photo-penguin coupling. 
TMM is defined as $\mu_{ij} = 2 i \Im F_2(q^2)_{ji}$. 
Note that both of lepton flavor and CP violations 
 are necessary for non-zero TMM to be induced
 (except the CP invariant case with relative CP-phase $\pi$). 
For Majorana neutrinos in the standard model 
 extended to incorporate the see-saw mechanism \cite{yanagida}, 
 TMM is found to be \cite{Fukugita-Yanagida} 
\bea
 \mu_{ij}=\frac{3 e G_F}{16 \pi^2 \sqrt{2}}(m_i+m_j)
 \sum_{\alpha =e, \mu, \tau}
 \Im \left[(U_{MNS}^{\dagger})_{j \alpha} 
 \left(\frac{m_\alpha}{M_W} \right)^2
 (U_{MNS})_{\alpha i}  \right]  \; ,  
 \label{SM}
\eea
where $m_i$ is the mass eigenvalue of the i-th generation neutrino, 
 $U_{MNS}$ is the Maki-Nakagawa-Sakata mixing matrix \cite{maki}, 
 and $M_W$ is the weak boson mass. 
Note that in order to evaluate the TMM 
 we need informations of the absolute values 
 of neutrino masses and all the elements in $U_{MNS}$ 
 including three CP-phases. 

If the model is supersymmetric (SUSY) one, 
 there are additional contributions to neutrino magnetic moments 
 through radiative corrections with sparticles running in a loop. 
It is worth investigating how large SUSY contributions can be. 
In minimal supersymmetric standard model (MSSM) 
 with Dirac neutrinos, 
 NMMs have been calculated and 
 found to be the same order of magnitude 
 as those in the standard model with Dirac neutrinos \cite{Ng}. 

In this article, we discuss the case of the MSSM 
 with Majorana neutrinos through the see-saw mechanism. 
As in the case of the standard model with Majorana neutrinos 
 discussed above, informations of neutrino masses and all the elements 
 of MNS mixing matrix are necessary for the evaluation of neutrino TMM. 
In addition, we need further informations 
 of neutrino Dirac Yukawa couplings and 
 the heavy right-handed neutrino masses, 
 since soft SUSY breaking parameters at the weak scale 
 depend on them through renormalization group equation (RGE) 
 evolutions. 
In the following, we evaluate the TMM 
 by using unambiguously determined neutrino mass matrices 
 in recently proposed minimal SUSY SO(10) model 
 \cite{Fukuyama-Okada}. 

Let us first give the general formula of SUSY contributions to TMM%
\footnote{ 
 The formulas with the mass insertion approximation 
 have been given in Ref.~\cite{Causse}.}.  
The basic theory which we analyze 
 is the MSSM with the right-handed Majorana neutrinos. 
The superpotential in the leptonic sector is given by 
\bea 
 W_Y =  Y_{\nu}^{ij} (\nu_R^c)_i \ell_j H_u 
 + Y_e^{ij} (e_R^c)_i \ell_j H_d 
 + \frac{1}{2} M_{R_{ij}} (\nu_R^c)_i  (\nu_R^c)_j 
   + \mu H_d  H_u  \;  , 
 \label{superpotential} 
\end{eqnarray} 
where the indeces $i$, $j$ run over three generations, 
 $H_u$ and $H_d$ denote the up-type and down-type MSSM 
 Higgs doublets, respectively, and  $M_R$ is 
 the heavy right-handed Majorana neutrino mass matrix.  
$Y_{\nu}$ and $Y_e$ are the neutrino Dirac Yukawa matrix  
 and the charged-lepton Yukawa matrix, respectively. 
In the following, we work in the basis where 
 $Y_e$ and $M_R$ are real-positive and diagonal matrices:  
 $Y_e^{ij}=Y_{e_i} \delta_{ij}$ and  
 $M_{R_{ij}}=\mbox{diag} ( M_{R_1},  M_{R_2},  M_{R_3}) $. 
The soft SUSY breaking terms in the leptonic sector 
 is described as 
\bea
-{\cal L}_{\mbox{soft}} &=& 
   \tilde{\ell}^{\dagger}_i 
   \left( m^2_{\tilde{\ell}} \right)_{ij}
   \tilde{\ell}_j 
 + \tilde{\nu}_{R i}^{\dagger} 
   \left( m^2_{ \tilde{\nu}} \right)_{ij}
    \tilde{\nu}_{R j} 
 + \tilde{e}_{R i}^{\dagger} 
   \left( m^2_{ \tilde{e}} \right)_{ij} 
\tilde{e}_{R j}    \nonumber  \\ 
&+& m_{H_u}^2 H_u^{\dagger} H_u + m_{H_d}^2 H_d^{\dagger} H_d  
+ \left(  B \mu H_d H_u 
+ \frac{1}{2} B_{\nu i j} \tilde{\nu}_{R i}^{\dagger} \tilde{\nu}_{R j} 
+ h.c. \right)  \nonumber \\ 
&+& \left( 
  A_{\nu}^{ij} \tilde{\nu}_{R i}^{\dagger}  \tilde{\ell}_j H_u 
+ A_e^{ij} \tilde{e}_{R i}^{\dagger} \tilde{\ell}_j H_d  +h.c.  
 \right)  \nonumber \\ 
&+& \left( 
    \frac{1}{2} M_1 \tilde{B}  \tilde{B}  
 +  \frac{1}{2} M_2 \tilde{W}^a  \tilde{W}^a  
  + \frac{1}{2} M_3 \tilde{G}^a  \tilde{G}^a  + h.c. \right)  \; .
 \label{softterms} 
\eea

At energies lower than $M_{R_i}$, 
 the right-handed neutrinos are decoupled, 
 and light Majorana neutrinos via the see-saw mechanism 
 are only neutrinos relevant for the low energy effective action. 
The effective interaction Lagrangian among the mass eigenstates of 
 light Majorana neutrino $\nu_i$, chargino $\chi_A^{-}$ 
 and charged-slepton $\tilde{e}_X$   
 is given by \cite{hisano} \cite{gunion} 
\bea
 {\mathcal L}_{int}= 
 \overline{\nu_i}\left( C^{R}_{iAX}P_R \right) 
 \tilde{\chi}^{+}_{A} \tilde{e}_{X} + h.c.
\label{interaction}
\eea
where $ P_{R} =  (1 + \gamma _5)/2 $ 
 is the right-handed chirality projection operator, and 
\bea
 C_{iAX}^R = g_2 (U_{MNS}^{*})_{\alpha i} 
\left( 
 - (O_L^*)_{A 1} (U_{\tilde e}^*)_{X \alpha} 
 + \frac{m_{\alpha}}{\sqrt{2}M_{\mathrm{W}}\cos \beta}
 (O_{L}^*)_{A 2} (U_{\tilde e}^*)_{X \alpha +3}  \right)   
\eea
with $m_\alpha=m_e, m_\mu, m_\tau$. 
Chargino mass matrix $M_C$ 
 and charged-slepton mass-squared matrix $M_{\tilde{e}}$ 
 are diagonalized
 by the unitary matrices $O_L~, ~ O_R$ and $U_{\tilde e}$, respectively, 
 such that 
\bea
 O_R M_c O_L^{\dagger} &=&\mbox{diag} 
 (m_{\tilde{\chi}_{1}^{-}}, m_{\tilde{\chi}_{2}^{-}})\; ,  
\nonumber \\
 U_{\tilde e}M_{\tilde{e}} U_{\tilde e}^\dagger &=&
 \mbox{diag}(m_{\tilde{e}_{1}}^2,....,m_{\tilde{e}_{6}}^2)\ ; .
\eea
Note that the left-handed chirality part is absent 
 in Eq.~(\ref{interaction})
 because of the decoupling of the right-handed Majorana neutrinos. 

One-loop corrections with chargino and charged-slepton running 
 in the loop give dominant contributions to the TMM. 
Corresponding Feynman diagrams are depicted in Fig.~1.  
Note that two diagrams are summed because of 
 Majorana property of the external line neutrinos. 
Considering all the diagrams, we obtain the TMM 
 normalized by the Bohr magneton such as 
\bea
\mu_{ij}
=- \frac{1}{16\pi^2}
 \sum_{A=1}^{2} \sum_{X=1}^{6}
 \frac{m_{e}}{m_{\tilde{\chi}_{A}^{-}}}
 \left[\frac{m_i+m_j}
 {m_{\tilde{\chi}_{A}^{-}}}
 \left( \Im[C_{iAX}^{R} C_{jAX}^{R*}] \right) \,f(x_{AX}) 
 \right] \mu_B .
 \label{TMM}
\eea
where $m_i$ is the neutrino mass of the i-th generation, 
 $x_{AX} = m^2_{\tilde{\chi}_{A}^{-}} /m^2_{\tilde{e}_X}$, 
 and $f(x)$ is the loop function defined as   
\bea
f(x) = \frac{x}{2(1-x)^2}
\left[\frac{2x}{1-x} \log x +1+x \right] \; ,
\eea
which is a monotonically increasing function of $x$ 
 varying from $ f(x \rightarrow +0)=0$ 
 to $f(x \rightarrow \infty)=\frac{1}{2}$. 
We can easily check that $\mu_{ij}=-\mu_{ji}$, and hence $\mu_{ii}=0$. 

As can be seen in the general formula of Eq.~(\ref{TMM}), 
 we need knowledge of neutrino mass eigenvalues, 
 all the components of the MNS mixing matrix 
 and the soft SUSY breaking terms at the weak scale, 
 for the evaluation of concrete TMM values. 
There are little models which can determine 
 all the fermion mass matrices including CP phases. 
In this article, we refer the results 
 in the recently proposed minimal SUSY SO(10) model 
 \cite{Fukuyama-Okada}, 
 where all the fermion mass matrices are unambiguously 
 determined. 
Using the results in \cite{Fukuyama-Okada}, 
 SUSY contributions to the lepton flavor violating processes 
 and muon $g-2$ has been calculated 
 based on the minimal supergravity (mSUGRA) boundary conditions 
 \cite{Fukuyama-etal}. 
In the following calculations, 
 we apply the same strategy and input parameters 
 as in \cite{Fukuyama-etal}. 

First we list the results in \cite{Fukuyama-Okada} 
 for the case of $\tan \beta =45$ 
 which are used in the following analysis. 
The light Majorana neutrino mass eigenvalues are fixed as (in GeV) 
 $m_1 = 2.45 \times 10^{-12}$, 
 $m_2 = 1.95 \times 10^{-11}$ and 
 $m_3 = 4.88 \times 10^{-11}$, 
 while the right-handed heavy Majorana neutrino mass eigenvalues 
 are found to be (in GeV) 
 $M_{R_1}=1.64 \times 10^{11}$,  
 $M_{R_2}=2.50 \times 10^{12}$ and 
 $M_{R_3}=8.22 \times 10^{12}$,  
 when a model parameter is fixed 
 so as to provide $\Delta m_\oplus^2 = 2 \times 10^{-3} \mbox{eV}^2$ 
 for the atmospheric neutrino oscillation data. 
The MNS mixing matrix has been found to be 
\begin{eqnarray}
 U_{MNS} = \left( 
 \begin{array}{ccc}
0.168 + 0.838 i & -0.467+0.0940 i & -0.00508+0.207 i \\
0.0519 + 0.498 i & 0.651-0.0473 i & 0.0189 -0.569 i \\ 
0.0745 + 0.116 i & 0.450-0.381 i & 0.431 + 0.669 i 
   \end{array} 
  \right) \; .  
\end{eqnarray}
In the basis where both of the charged-lepton 
 and right-handed Majorana neutrino mass matrices 
 are diagonal with real and positive eigenvalues, 
 the neutrino Dirac Yukawa matrix at the grand unification (GUT) scale 
 is found to be 
\begin{eqnarray}
 Y_{\nu} = 
\left( 
 \begin{array}{ccc}
-0.000135 - 0.00273 i & 0.00113  + 0.0136 i  & 0.0339   + 0.0580 i  \\ 
 0.00759  + 0.0119 i  & -0.0270   - 0.00419  i  & -0.272    - 0.175   i  \\ 
-0.0280   + 0.00397 i & 0.0635   - 0.0119 i  &  0.491  - 0.526 i 
 \end{array}   \right) \; .  
\label{Ynu}
\end{eqnarray}     

The soft SUSY breaking parameters at the weak scale 
 are evaluated through the RGE evolutions 
 by imposing mSUGRA universal boundary conditions 
 at the GUT scale, $M_{GUT} \sim 2 \times 10^{16}$GeV, 
 such that 
\begin{eqnarray}
 & & \left( m^2_{\tilde{\ell}} \right)_{ij} 
 = \left( m^2_{ \tilde{\nu}} \right)_{ij}
 =   \left( m^2_{ \tilde{e}} \right)_{ij} = m_0^2 \delta_{ij} \; , 
  \nonumber \\ 
 & &m_{H_u}^2=m_{H_d}^2 = m_0^2  \; , 
  \nonumber \\ 
 & & A_{\nu}^{ij} = A_0 Y_{\nu}^{ij}\; , \; \; 
  A_{e}^{ij} = A_0 Y_{e}^{ij} \; , 
  \nonumber \\ 
 & & M_1=M_2=M_3= M_{1/2} \; .  
 \label{mSUGRA}
\end{eqnarray}
The neutrino Dirac Yukawa matrix contributes 
 the RGE evolutions of the soft SUSY breaking parameters 
 at the energy scales above $M_{R_i}$. 
For example, RGE for the left-handed slepton mass-squared matrix 
 is given by 
\begin{eqnarray}
 \mu \frac{d}{d \mu} 
  \left( m^2_{\tilde{\ell}} \right)_{ij}
 &=&  \mu \frac{d}{d \mu} 
  \left( m^2_{\tilde{\ell}} \right)_{ij} \Big|_{\mbox{MSSM}} 
  \nonumber \\
 &+& \frac{1}{16 \pi^2} 
 \left( m^2_{\tilde{\ell}} Y_{\nu}^{\dagger} Y_{\nu}
  + Y_{\nu}^{\dagger} Y_{\nu} m^2_{\tilde{\ell}} 
  + 2  Y_{\nu}^{\dagger} m^2_{\tilde{\nu}} Y_{\nu}
  + 2 m_{H_u}^2 Y_{\nu}^{\dagger} Y_{\nu} 
  + 2  A_{\nu}^{\dagger} A_{\nu} \right)_{ij} , 
 \nonumber\\ \; 
\label{RGE} 
\end{eqnarray}
where the first term in the right hand side denotes 
 the normal MSSM term. 
In the leading-logarithmic approximation, 
 additional contribution due to the neutrino Dirac Yukawa couplings   
 are estimated as 
\begin{eqnarray}
 \left(\Delta  m^2_{\tilde{\ell}} \right)_{ij}
 \sim - \frac{3 m_0^2 + A_0^2}{8 \pi^2} 
 \left( Y_{\nu}^{\dagger} L Y_{\nu} \right)_{ij} \; ,  
 \label{leading}
\end{eqnarray}
where the distinct thresholds of the right-handed 
 Majorana neutrinos are taken into account 
 by the matrix $ L_{ij} = \log [M_G/M_{R_i}] \delta_{ij}$.  
 
Now let us present our numerical results.  
The resultant $|\mu_{ij}|$ as a function of 
 the universal scalar mass $m_0$ 
 is depicted in Fig.~2 
 for fixed  $M_{1/2}=600$GeV and $A_0=0$ 
 together with the results in the standard model 
 with the see-saw mechanism in Eq.~(\ref{SM}). 
The resultant magnetic moments are decreasing 
 as $m_0$ becomes large, according to Eq.(15) with 
 $X_{AX}  \sim  M_{1/2}^2/m_0^2$ for $m_0 > M_{1/2}$. 
We find that the SUSY contributions to TMM 
 with the input soft SUSY breaking masses being of order 1 TeV 
 can be an order of magnitude larger than those in the standard model. 
In Fig.~3, $|\mu_{ij}|$ as a function of 
 the universal gaugino mass $M_{1/2}$ 
 is depicted for fixed  $m_0=400$GeV and $A_0=0$. 
We can read off the fact that 
 the resultant TMM is roughly proportional to $M_{1/2}^{-2}$.  
For fixed $m_0=600$GeV and $M_{1/2}=800$GeV, 
 TMM is plotted in Fig.~4 as a function of $A_0$. 
It is found that TMM is not so sensitive to $A_0$. 

In conclusion, we have examined 
 the SUSY contributions to TMM of Majorana neutrinos.  
In concrete evaluations, 
 informations of neutrino mass matrix elements  
 including CP phases are necessary. 
Using unambiguously determined neutrino mass matrices 
 in the minimal SUSY SO(10) model in \cite{Fukuyama-Okada}, 
 TMM has been calculated. 
We have found that
 the SUSY contributions to TMM 
 with the input soft SUSY breaking masses being of order 1 TeV 
 can be an order of magnitude larger 
 than those calculated in the standard model 
 extended to incorporate the see-saw mechanism. 
Unfortunately, the calculated TMMs are found to be too small 
 to cause interesting astrophysical phenomena. 
This fact can be understood in intuitive way 
 with a rough estimation of TMM. 
We consider neutrinos as the Majorana particles 
 via the see-saw mechanism with heavy right-handed Majorana neutrinos.
At weak scale, the right-handed neutrinos are decoupled, 
 and only the light Majorana neutrinos appear 
 in low energy effective theory 
 and only their left-handed components can couple 
 to SUSY particles as in Eq.~(\ref{interaction}). 
In this case, the chirality flip between in and out neutrino states
 necessary for TMM 
 can occur only by mass insertions in the external lines. 
Therefore, TMM is always proportional to 
 the light Majorana masses. 
Taking this fact into account, 
 we can estimate the order of magnitude of 
 the SUSY contributions such as 
\bea
\mu_\nu \sim e\times 10^{-2} \frac{m_\nu}{M_{SUSY}^2} 
 \sim 10^{-18} 
 \left( \frac{m_\nu}{\mathrm{1eV}} \right) 
 \left( \frac{\mathrm{100GeV}}{M_{SUSY}} \right)^2 \; \mu_B
\eea
where $e$ is the electric charge, 
 $10^{-2}$ is a loop factor, 
 and $m_\nu$ and $M_{SUSY}$ are 
 neutrino mass and typical soft SUSY breaking mass, respectively.
We can see that this formula gives a good approximation. 

%

%
\newpage
\begin{figure}
\begin{center}
\epsfig{file=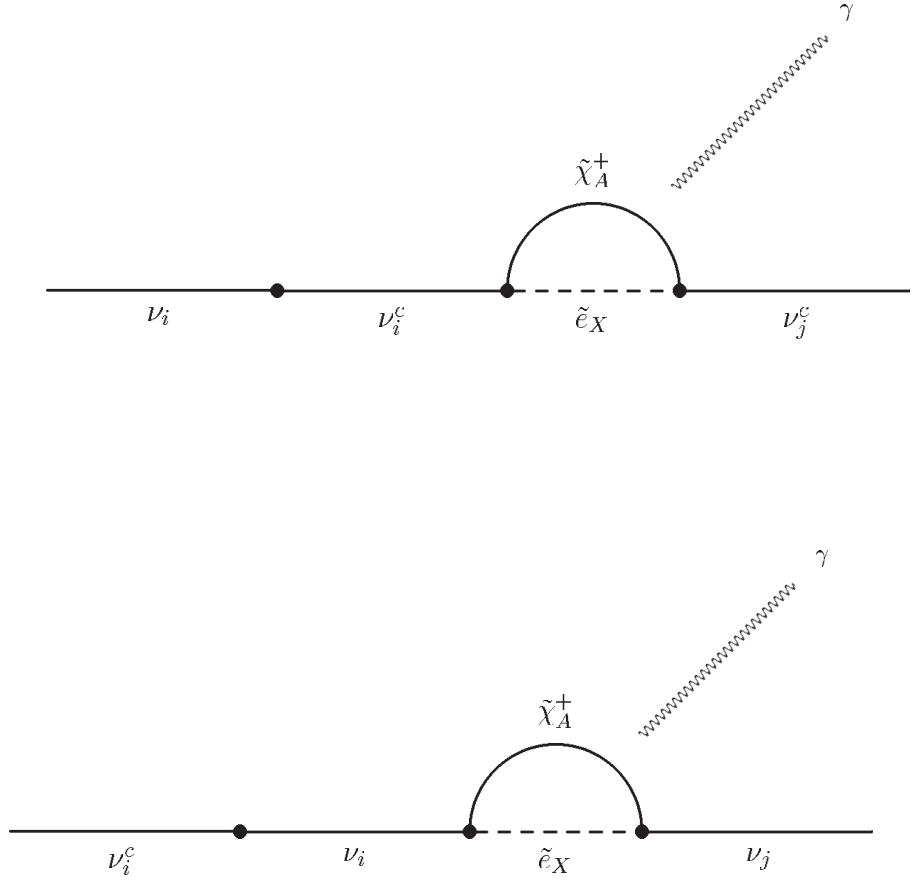, width=12cm}
\caption{
The 1-loop Feynman diagrams which generate 
 the transition magnetic moments of the Majorana neutrinos.  
Because of the Majorana property of the external neutrino, 
 we have to sum contributions of two diagrams.}
\end{center}
\end{figure}
\begin{figure}
\begin{center}
\epsfig{file=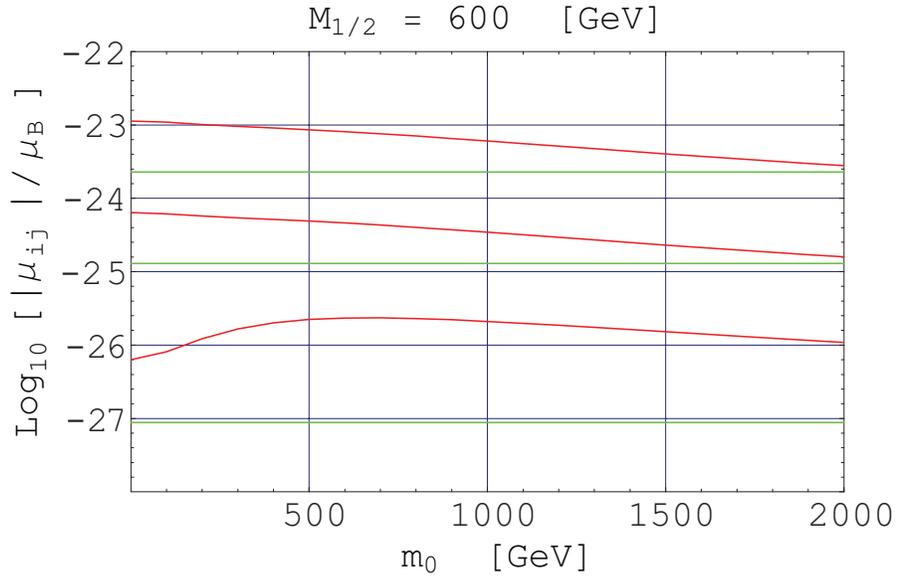, width=12cm}
\caption{
The transition magnetic moments of the Majorana neutrinos,  
$ \mbox{Log}_{10}[|\mu_{23}|] $,
$ \mbox{Log}_{10}[|\mu_{12}|] $ and 
$ \mbox{Log}_{10}[|\mu_{13}|] $ 
from top to bottom as a function of $m_0$ (GeV) 
with fixed $M_{1/2}=600$GeV and $A_0=0$. 
The horizontal lines denote the results 
 from the standard model with see-saw mechanism 
 with the same MNS matrix , 
 and the lines correspond to 
 $ \mbox{Log}_{10}[|\mu_{23}|]$,
 $ \mbox{Log}_{10}[|\mu_{12}|]$ and 
 $ \mbox{Log}_{10}[|\mu_{13}|]$ 
 from top to bottom, respectively. 
}
\end{center}
\end{figure}
\begin{figure}
\begin{center}
\epsfig{file=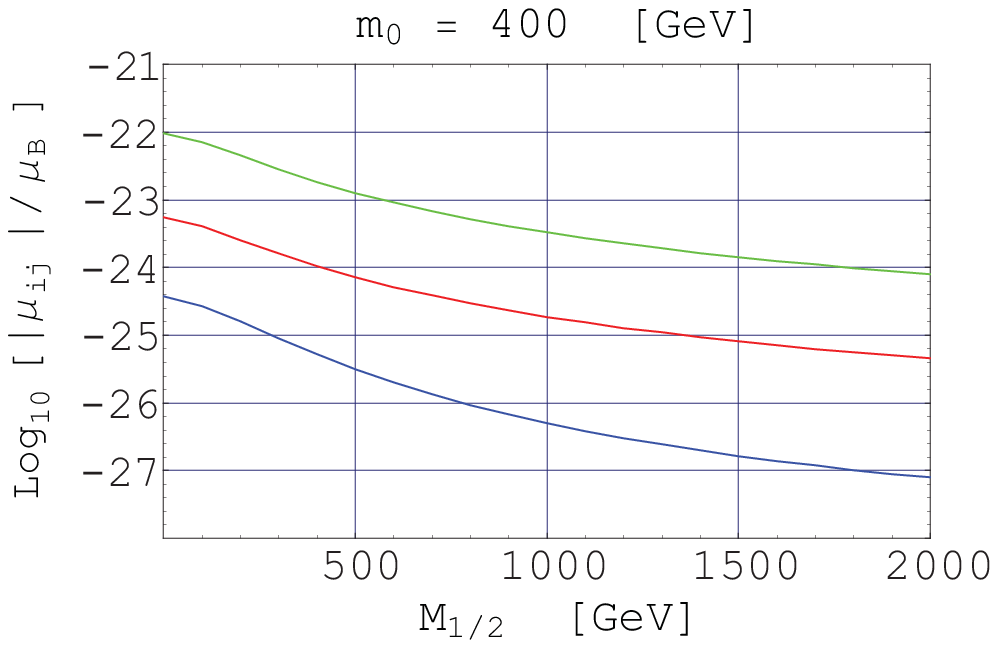, width=12cm}
\caption{
The transition magnetic moments of the Majorana neutrinos,  
$ \mbox{Log}_{10}[|\mu_{23}|] $,
$ \mbox{Log}_{10}[|\mu_{12}|] $ and 
$ \mbox{Log}_{10}[|\mu_{13}|] $ 
from top to bottom as a function of $M_{1/2}$ (GeV) 
 with fixed $m_0=400$GeV and $A_0=0$. 
}
\end{center}
\end{figure}
\begin{figure}
\begin{center}
\epsfig{file=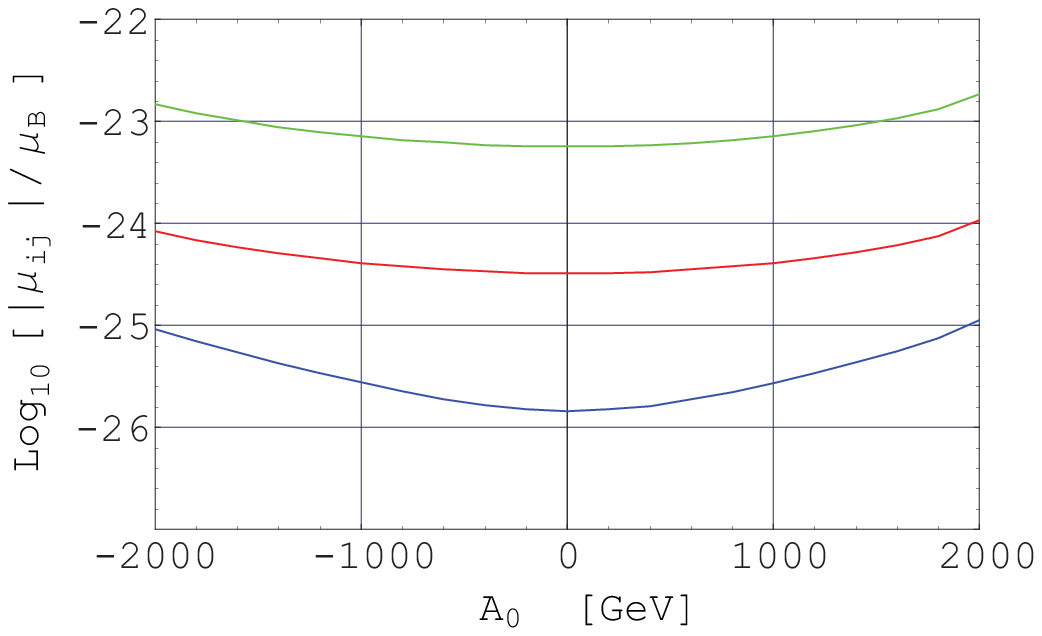, width=12cm}
\caption{
The transition magnetic moments of the Majorana neutrinos,  
$ \mbox{Log}_{10}[|\mu_{23}|] $,
$ \mbox{Log}_{10}[|\mu_{12}|] $ and 
$ \mbox{Log}_{10}[|\mu_{13}|] $ 
from top to bottom as a function of $A_0$ (GeV) 
 with fixed $m_0=600$GeV and $M_{1/2}=800$GeV. 
}
\end{center}
\end{figure}

\begin{thebibliography}{99}
%
\bibitem{PDG}
K.~Hagiwara {\it et al.}  [Particle Data Group Collaboration], 
Phys. Rev. {\bf D66}, 010001 (2002).
\bibitem{raffelt}
M. Fukugita and S. Yazaki,  Phys. Rev. {\bf D36}, 3817 (1987);
M. Haft, G.G. Raffelt, and A. Wiess, Astrophys. J. {\bf 425}, 222 (1994). 
\bibitem{ayala}
A. Ayala, J.C. D'Olivo, and M. Torres, Phys. Rev. {\bf D59}, 111901 (1999)
\bibitem{lujan}
E. Torrente-Lujan, JHEP {\bf 0304}, 054, (2003). 
\bibitem{akhmedov1}
E.Kh. Akhmedov and Joao Pulido, Phys. Lett. 
{\bf B553}, 7 (2003); 
E.Kh. Akhmedov and T. Fukuyama, JCAP {\bf 0312}, 007 (2003). 
\bibitem{akhmedov2}
H. Athar, J.T. Peltoniemi, and A.Yu. Smirnov, 
Phys. Rev. {\bf D51}, 6647 (1995); 
T. Totani and K. Sato, Phys. Rev. {\bf D54}, 5975 (1996);~
E.Kh. Akhmedov, C. Lunardini, and 
A.Yu. Smirnov, Nucl. Phys. {\bf B643}, 339 (2002);~
S. Ando and K. Sato, Phys. Rev. {\bf D67}, 023004 (2003).
\bibitem{raffelt1}
G.G. Raffelt, ``Stars as Laboratories 
 for Fundamental Physics" (The University of Chicago Press ;1996).
\bibitem{fujikawa}
K. Fujikawa and R.E. Shrock, Phys. Rev. Lett. 
{\bf 45}, 963 (1980).
\bibitem{yanagida}
T. Yanagida, in Proceedings of the workshop 
 on the Unified Theory and Baryon Number in the Universe, 
 edited by O.Sawada and A.Sugamoto (KEK, Tsukuba, 1979);
M. Gell-Mann, P. Ramond, and R. Slansky, 
 in Supergravity, edited by D.Freedman and P.van Niewenhuizen 
 (north-Holland, Amsterdam 1979); 
R.~N.~Mohapatra and G.~Senjanovic,
Phys.\ Rev.\ Lett.\  {\bf 44}, 912 (1980).
\bibitem{Fukugita-Yanagida}
M. Fukugita and T. Yanagida,
``Physics of neutrinos'' in ``Physics and Astrophysics of Neutrinos" ;
 A. Fukugita and A. Suzuki eds. (Springer-Verlag ; 1994).
\bibitem{maki}
J. Maki, M. Nakagawa and S. Sakata, 
Prog. Theor. Phys. 28, 870 (1962). 
\bibitem{Ng} 
K.L. Ng, Z. Phys. C48, 289 (1990), 
 and see the references for the previous works. 
\bibitem{Fukuyama-Okada} 
T. Fukuyama and N. Okada, JHEP {\bf 0211}, 011 (2002).
\bibitem{Causse}
M.B. Causse, arXiv:hep-ph/0202096.
\bibitem{hisano}
J. Hisano, T.Moroi, K. Tobe, and M. Yamaguchi, Phys. Rev. 
{\bf D53}, 2442 (1996).
\bibitem{gunion}
J. Gunion and H. Harber, Nucl. Phys. {\bf B272}, 1 (1986); 
{\bf B278}, 449 (1986).
\bibitem{Fukuyama-etal}
T.~Fukuyama, T.~Kikuchi and N.~Okada, 
Phys. Rev. {\bf D68}, 033012, (2003).
%
\end{thebibliography}
\end{document}